\documentclass[aps,prc,preprintnumbers,showpacs,superscriptaddress,twocolumn,floatfix]{revtex4-1}
\usepackage{graphicx}
\usepackage{dcolumn}
\usepackage{bm}
\usepackage{epsf}
\usepackage{epsfig}
\usepackage{amssymb}
\usepackage{verbatim}
\usepackage{amsmath}
\usepackage{longtable}
\usepackage[utf8]{inputenc}
\usepackage{helvet}
\usepackage[polish,english]{}
\usepackage{subdepth}
\usepackage{tikz}
\usepackage{rotating}
\usepackage{footnote}
\usepackage{threeparttable}
\usepackage{color}
\usepackage{chngpage}
\usepackage{array}
\usepackage{booktabs}
\usepackage{tabularx}
\usepackage{lipsum}
\begin{document}
\title{The anomalous quadrupole collectivity in Te isotopes}
\author{Chong~Qi}
\email{chongq@kth.se}
\affiliation{Department of Physics, KTH Royal Institute of Technology, 10691 Stockholm, Sweden}
\date{\today}
\begin{abstract}
We present systematic calculations on the spectroscopy and transition properties of even-even Te isotopes by using the large-scale configuration interaction shell model approach with a realistic interaction.
These nuclei are of particular interest since their yrast spectra show a vibrational-like equally-spaced pattern but the few known E2 transitions show anomalous rotational-like behavior, which cannot be reproduced by collective models. Our calculations reproduce well the equally-spaced spectra of those isotopes as well as the constant behavior of the $B(E2)$ values in $^{114}$Te. The calculated  $B(E2)$ values for neutron-deficient and heavier Te isotopes show contrasting different behaviors along the yrast line. The $B(E2)$ of light isotopes can exhibit a nearly constant bevavior upto high spins. We show that this is related to the enhanced neutron-proton correlation when approaching $N=50$. 

\end{abstract} 
\maketitle
The advent of large-scale radioactive beam facilities and new detection technologies have enabled the study of the spectroscopy and transition properties of $N\sim Z$ nuclei just above the presumed doubly magic nucleus $^{100}$Sn \cite{Faestermann201385}. Several unexpected phenomena have recently been observed: Large $B(E2;2^+_1\rightarrow0_1^+)$ values of neutron deficient semi-magic $\mathrm{Sn}$ isotopes have triggered extensive experimental~\cite{PhysRevC.72.061305,Corsi2015451,PhysRevLett.98.172501,PhysRevC.93.021601,PhysRevLett.101.012502,PhysRevC.90.061302,Jungclaus2011110,PhysRevLett.110.172501,PhysRevLett.101.012502,PhysRevC.92.041303,Spieker2016102,PhysRevC.92.014330}
and theoretical \cite{Ansari2007128,PhysRevC.84.044314,Morales2011606,PhysRevC.89.014320,PhysRevC.86.054304,PhysRevC.91.041301,Engeland201451,Back2013,PhysRevC.86.044323} activities, in particular regarding the fundamental roles played by core excitations and the nuclear pairing correlation (or seniority coupling). 
The study of transition rates in isotopic chains just above $Z=50$ may provide further information on the role of core excitations \cite{PhysRevC.84.041306,PhysRevC.91.061304}. 
The limited number of valence protons and neutrons are not expected to induce any significant quadrupole correlation in this region \cite{Hadinia2005,PhysRevC.82.064304,PhysRevC.92.024306,PhysRevC.92.064309}.
The low-lying collective excitations of those nuclei were discussed in terms of quadrupole vibrations \cite{Hadinia2005,PhysRevC.51.2394} in relation to the fact that the even-even Te isotopes between $N=56$ and 70 show regular equally-spaced yrast spectra (c.f., Fig. 1 in Ref. \cite{Hadinia2005}). If that is the case, the Te isotopes will provide an ideal ground to explore the nature of the elusive nuclear vibration and the residual interactions that leading to that collectivity. However, the available E2 transition strengths along the yrast line in $^{114,120-124}$Te show an anomalous rotational-like behavior, which can not be reproduced by collective models or the interacting boson model \cite{PhysRevC.71.064324,PhysRevC.90.024316}. 
Another intriguing phenomenon is the nearly constant behavior of the energies of the $2^+$ and 
$4^+$ states in Te and Xe isotopes and their ratios when approaching $N=50$, in contrast to the decreasing behavior when approaching $N=82$ \cite{Hadinia2005}. This was analyzed in Ref. \cite{Delion2010} based on the quasiparticle random phase approximation approach where an competition between the quadurople-quadrupole correlation and neutron-proton pairing correlation was suggested.

An enhanced interplay between neutrons and protons is expected in the $^{100}$Sn region since the protons and neutrons partially occupy the same quantum orbitals near the Fermi level \cite{Hadinia2004,Hadinia2005,Sandzelius2007,Delion2010}. In relation to that, there has also been a long effort searching for superallowed alpha decays from those $N\sim Z$ isotopes \cite{Seweryniak2006,Liddick2006}.
The region is also expected to be the endpoint of the astrophysical
rapid proton capture (rp) process \cite{PhysRevLett.86.3471,PhysRevLett.102.252501}. The octupole correlation may also play a role here (the coupling between the $0h_{11/2}$
and $1d_{5/2}$ orbitals) \cite{PhysRevC.61.064305,ISI:000259052900005,PhysRevC.92.024306}.
Still, compared to tin, the experimental information is less abundant in the isotopic chain of tellurium where little was known experimentally below the neutron midshell until recently. Much more work is needed and further measurements are underway in order to map out the ordering and nature of single-particle states and two-body effective interactions in the region \cite{Cederwall2016,Grahn:1981273}. 

In this work we present systematic large-scale calculations on the spectroscopy and transition properties of even-even Te isotopes. The large-scale shell model, which takes into account all degrees of freedom within a given model space, is an ideal approach to study competition between collective and single-particle degrees of freedom. 
It is however a challenge, especially in the midshell, due to the huge dimension of the problem (c.f., Fig. 1 in Ref. \cite{PhysRevC.86.044323}). 
On the theoretical side,
we have done systematic calculations on the $B(E2;2^+_1\rightarrow0^+_1)$ values of even-even Te isotopes \cite{PhysRevC.84.041306,PhysRevC.91.061304}. The results are, however, rather sensitive to the truncation imposed. Now we are able to do full shell model calculations for all low-lying states of all Te isotopes with further optimization of the shell-model algorithm. 
A full shell model calculation for the spectroscopy of $^{104}$Te was done in Ref. \cite{Faestermann201385}. A schematic calculation for $^{106}$Te in the $1d_{5/2}0g_{7/2}$ subspace was presented in Ref. \cite{Hadinia2005}. 
Systematic calculations on the Sn and Sb isotopes were given in Refs. \cite{,Back2013,PhysRevC.86.044323} and Ref. \cite{Honma2014}, respectively. In Ref. \cite{PhysRevC.92.064309}, the possible onset of vibrational collectivity in Te isotopes was discussed within an effective field theory framework.

\begin{figure*}[htp]
\begin{center}
\includegraphics[scale=0.45]{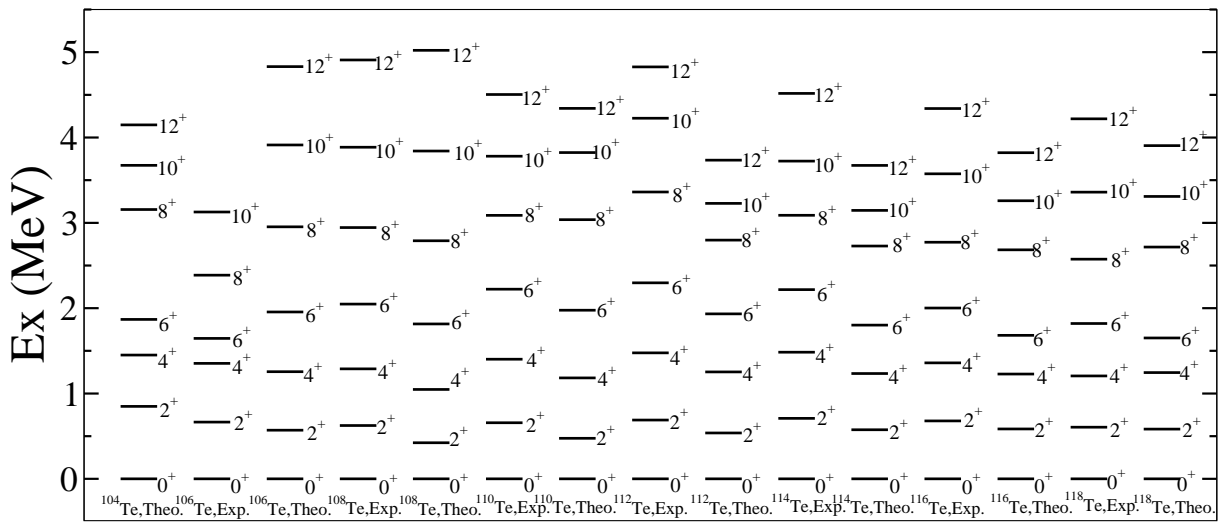}
\end{center}
\caption{\label{SM} Comparison between theory and experiment for the yrast spectra
of the Te isotopic chain.}
\end{figure*}


We consider the neutron and proton orbitals between the shell closures $N$ (and $Z$) $=50$ and 82, comprising $0g_{7/2}$, $1d_{5/2}$, $1d_{3/2}$, $2s_{1/2}$ and $0h_{11/2}$ and assume $^{100}$Sn as the inert core. The robustness of the $N=Z=50$ shell closures, which has fundamental influence on our understanding of the structure of nuclei in this region, is supported by recent measurements \cite{PhysRevC.84.041306,Back2013,10.1038/nature11116,PhysRevLett.110.172501,PhysRevC.91.061304}.
The nearly degenerate neutron single-particle states $d_{5/2}$ and $g_{7/2}$ orbitals in $^{101}$Sn
were
observed by studying the $\alpha$-decay $^{105}$Te $\rightarrow$ $^{101}$Sn \cite{PhysRevLett.97.082501,Seweryniak2006,PhysRevLett.99.022504,PhysRevLett.105.162502}. 
Based on the assumption that the ground state of $^{105}$Te has spin-parity $5/2^+$, the $g_{7/2}$ orbital was suggested to be the ground state of $^{101}$Sn instead of $d_{5/2}$.
The excitation energy of the $1d_{5/2}$ is taken as $\varepsilon (1d_{5/2})=0.172$ MeV. The energies of other states have not been measured yet. They are adjusted to fit the experimental binding energies of tin isotopes. 
The starting point of our calculation is the realistic CD-Bonn nucleon-nucleon potential \cite{Machleidt01}. The interaction was renormalized using the perturbative G-matrix approach to take into account the core-polarization effects \cite{hJensen95}. The $T=1$ part of the monopole interaction was optimized by fitting to the low-lying states in Sn isotopes \cite{PhysRevC.86.044323}. Further optimization of the $T=0$ part of the interaction is also underway which, however, is still a very challenging task. 
Our calculations show that the present effective Hamiltonian are already pretty successful in describing the structure and transition properties of Sb, Te, I, Xe and Cs isotopes as well as heavy nuclei 
near $N=82$
in this region.

The Te isotopic chain is the heaviest and longest chain that can be described by the nuclear shell model. The dimension for the mid-shell $^{118}$Te reaches $10^{10}$ for which the diagonalization is still a very challenging numeric task. In our previous calculations for the $B(E2;2^+_1\rightarrow0^+_1)$ of mid-shell Te isotopes~\cite{PhysRevC.84.041306}, we restricted a maximum of four neutrons that can be excited from below the Fermi surface to the neutron $h_{11/2}$ subshell and excluded proton excitation to $h_{11/2}$ due to limited computation power, which, as we understand now, is a rather severe truncation. The model space was further extended to allow at most 8 particles to the $h_{11/2}$ subshell in Ref. \cite{PhysRevC.91.061304}, where the $B(E2;2^+_1\rightarrow0^+_1)$ values show a much smoother parabolic behavior as a function of $N$.
Full shell-model calculations are done for all nuclei in the present work.
All shell-model calculations are carried out within the $M$-scheme where states with $M=I$ are considered. Diagonalizations are done with a parallel shell model program that we developed \cite{ISI:000261998300029} and with a slightly modified version of the code KSHELL \cite{Shi13}.
All calculations are done on the supercomputers Beskow and Tegn\'er at PDC Center for High Performance Computing at the KTH Royal Institute of Technology in Stockholm, Sweden. 


To test the validity of the effective interaction, we have done systematic calculations on the yrast spectra of isotopes $^{104-132}$Te. The results for $^{104-118}$Te are plotted in Fig. \ref{SM} in comparison with available experimental data~\cite{nudat}. An overall good agreement between theory and experiment is obtained. Noticeable difference is seen in the excitation energies of the $12^+$ states in $^{112,114}$Te and the $I\geq 6$ states in $^{106}$Te. All isotopes plotted in the figure show rather regular and vibrational-like spectra up to $12^+$ except $^{106}$Te. For that nucleus, the calculated spectrum still shows a equally-spaced pattern. However, a smaller gap between $6_1^+$ and $4^+_1$ states is expected from recent measurement but the spin-parity assignments for those states are still tentative. The equally-spaced pattern breaks down in isotopes heavier than $^{126}$Te where a gradual depression of the excitation energies of the $6^+$ states is seen.

\begin{figure}
\begin{center}\includegraphics[scale=0.9]{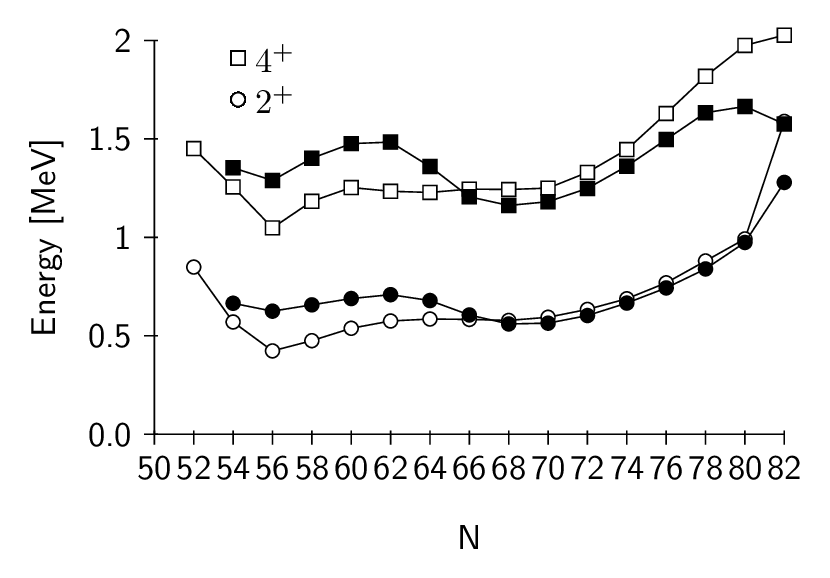}
\includegraphics[scale=0.9]{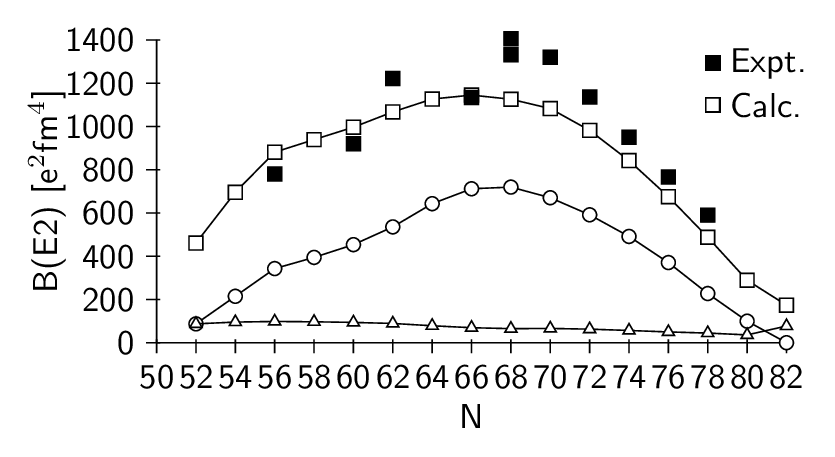}
\end{center}
\caption{\label{SM2} Comparison between theory and experiment for the energies of the $2^+$ and $4^+$ yrast states (upper) and the $B(E2;2^+_1\rightarrow0^+_1)$ values (lower) for the Te isotopic chain. The open circles and open triangles in the lower panel correspond to the square of the neutron and proton matrix elements, $M_n^2$ and $M_p^2$, respectively.}
\end{figure}

A closer comparison between experiment and calculation on the excitation energies of the yrast $2^+$ and $4^+$ states are plotted in Fig. \ref{SM2} as a function of neutron number for all even-even Te isotopes. 
In the lower panel of the figure, the calculated $B(E2;2^+_1\rightarrow0^+_1)$ in tellurium isotopes are compared to the most recent experimental data \cite{nudat,PhysRevC.84.041306,PhysRevC.91.061304}. The B(E2) value is calculated as $B(E2)$ = $(e_{p}M_{p}+e_{n}M_{n})^2$ where $M_p$ and $M_n$ are the proton and neutron matrix elements and we take effective charges $e_{p}$ = 1.5e and $e_{n}$ = 0.8e as were employed in~\cite{PhysRevC.84.041306,PhysRevC.91.061304}. The isospin dependence of the effective charges is not considered here, which is not expected to have large influence on the trend.
The model prediction agrees rather well with available data. 
The largest deviations are seen in isotopes $^{120,122}$Te. A recent measurement gave a value smaller than the adapted one in the former case.
In the figure we also plotted the square of the neutron and proton matrix elements, $M_n^2$ and $M_p^2$, which represent the separate contributions to the $B(E2)$ values from the neutron and proton excitations. 
As can be seen from the figure, the parabolic behavior of the $B(E2)$ values, which looks similar to that of Sn isotopes, is mostly due to the contribution from the neutron excitation. The contribution from the proton excitation shows a rather smooth and slightly decreasing behavior as the neutron number increases.
As mentioned earlier, the shell-model calculations for mid-shell Te isotopes, in particular $^{118}$Te, are quite sensitive to the filling of both the proton and neutron $h_{11/2}$ subshells. Both the proton and neutron transition matrix elements are enhanced when one goes from a small model space calculation with restricted number of particles in $h_{11/2}$ to the full shell-model calculation. 

\begin{figure}
\begin{center}
\includegraphics[scale=0.9]{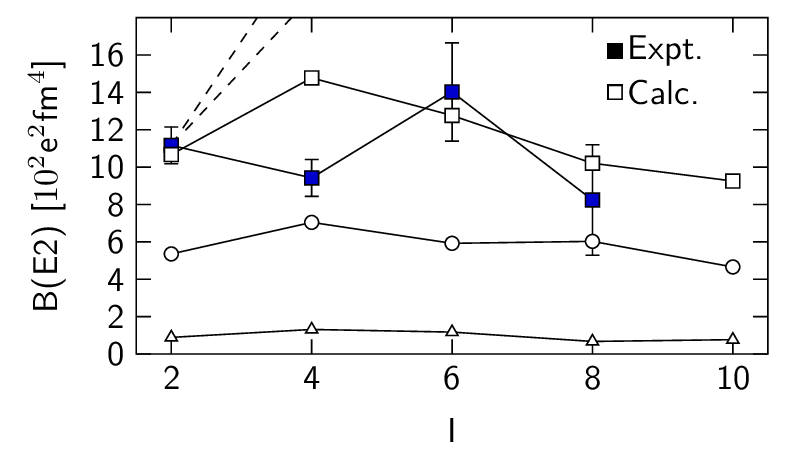}
\end{center}
\caption{\label{SM3} Comparison between theory and experiment \cite{PhysRevC.71.064324} for $B(E2;I^+_1\rightarrow(I-2)^+_1)$ values in $^{114}$Te along the yrast line. The open circles and open triangles in the lower panel correspond to the square of the neutron and proton matrix elements, $M_n^2$ and $M_p^2$, respectively. The dashed lines correspond to the predictions of collective models \cite{PhysRevC.71.064324}.}
\end{figure}

The regularly-spaced level spectra in mid-shell Te isotopes have been expected to be be associated with a collective vibrational motion. 
For a spectrum corresponds to a vibrator, there should be collective E2 transitions between states differing by one phonon. The transition strengths should be linearly proportional to the 
spin of the initial states, i.e., one has $B(E2,4_1^+\rightarrow2_1^+)/B(E2,2_1^+\rightarrow0_1^+)$=2 in the harmonic vibrator model.
Unfortunately, there are very few data available for $B(E2)$ values in states beyond $2^+_1$.
As shown in Ref. \cite{PhysRevC.71.064324}, the measured $B(E2)$ values along the yrast line in $^{114}$Te show a rather anomalous constant behavior up to $I=8$, which looks more like a rotor and is in contradiction with that for a vibrator. Our shell model calculations for those E2 transitions are shown in Fig. \ref{SM3}, which indeed exhibits a rather constant behavior up to higher spins. Moreover, both the proton and neutron matrix elements remain roughly the same along the yrast line. 

As can be seen Fig. \ref{SM3}, the ratio $B(E2,4_1^+\rightarrow2_1^+)/B(E2,2_1^+\rightarrow0_1^+)$ for $^{114}$Te is measured to be even slightly smaller than one. This is not seen in the theory. The ratio is calculated to be $B(E2,4_1^+\rightarrow2_1^+)/B(E2,2_1^+\rightarrow0_1^+)=1.38$ which actually agree well with the prediction for a rotor.
As discussed in Refs. \cite{PhysRevC.70.047302,PhysRevC.90.034307}, it happens rarely in open-shell nuclei that one has $B(E2,4_1^+\rightarrow2_1^+)/B(E2,2_1^+\rightarrow0_1^+)<1$. The reason why the ratio for   $^{114}$Te is observed to be so small  is still not clear.

\begin{figure}
\begin{center}
\includegraphics[scale=0.9]{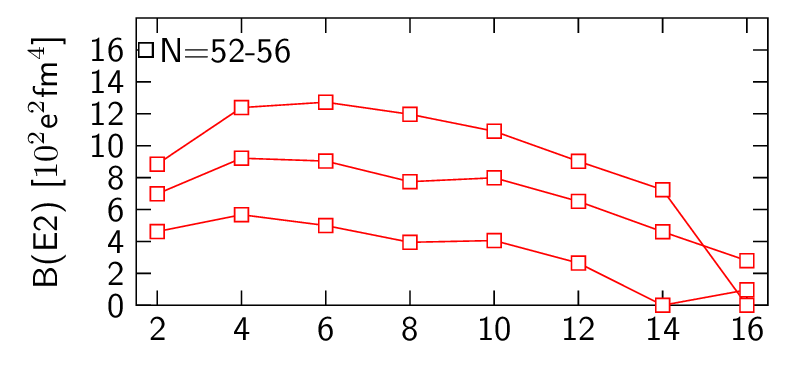}
\includegraphics[scale=0.9]{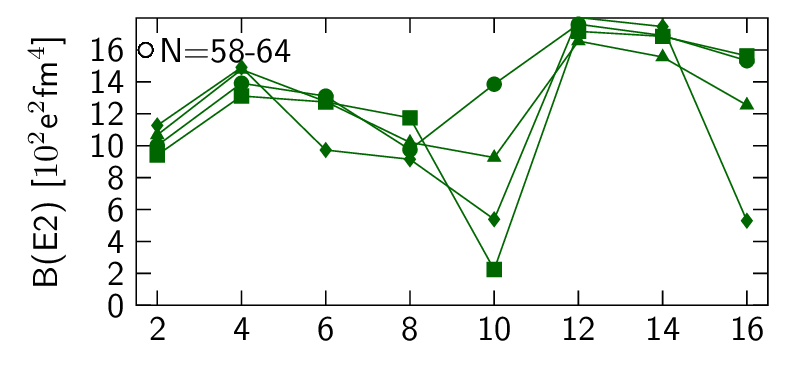}
\includegraphics[scale=0.9]{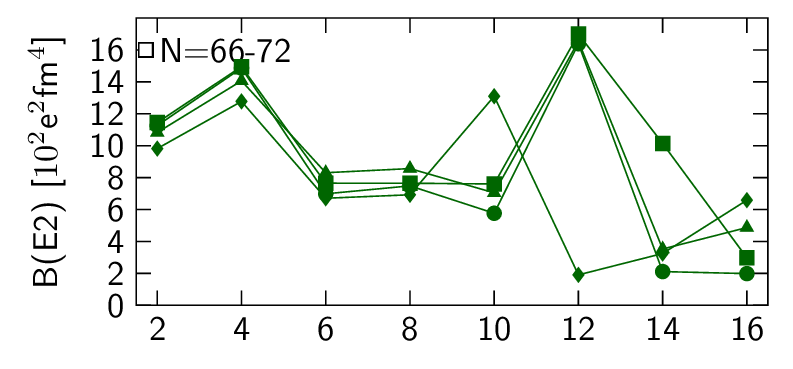}
\includegraphics[scale=0.9]{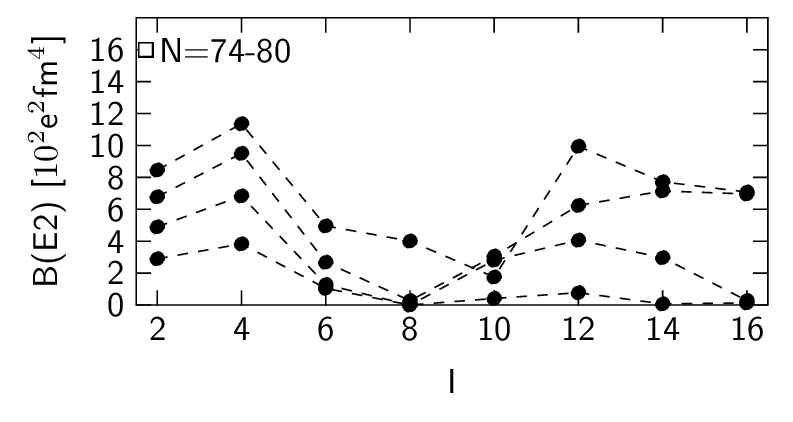}
\end{center}
\caption{\label{SM4} Calculated $B(E2;I^+_1\rightarrow(I-2)^+_1)$ values along the yrast line for even-even Te isotopes. }
\end{figure}

In Ref. \cite{PhysRevC.90.024316}, the ratios $B(E2,4_1^+\rightarrow2_1^+)/B(E2,2_1^+\rightarrow0_1^+)$ for isotopes $^{120-124}$Te are measured to be 1.640, 1.500 and 1.162, respectively.
The ratios calculated from the shell model $B(E2)$ values are 1.322, 1.299, and 1.301, respectively, for above three nuclei, which agree reasonably with experimental data.

In Fig. \ref{SM4} we plotted the calculated $B(E2)$ values for the yrast states of all even-even Te isotopes between $N=52$ and 80.
As can be seen from the upper panel of the figure, the $B(E2)$ values for the yrast states of $^{104-108}$Te, which are at the beginning of the shell, remain roughly constants up to spin $I=12$ and decrease significantly around $I=16$, which indicates that the collectivity has collapsed there. On the other hand, as shown in the lower panel of the figure, the results for the isotopes $^{126-132}$Te at the end of the shell show a very different behavior: The $B(E2)$ values decrease dramatically after $I=4$, which reach practically zero value for states up to $I=10$ in all nuclei except $^{128}$Te.
The results for the groups $N=58-64$ and $N=66-72$ are shown in the middle panels of Fig. \ref{SM4}. In the former group, the $B(E2)$ values show a rather constant behavior up to $I=8$. The results for $I=10$ diverge in relation to the fact that several low-lying $10^+$ states are predicted for those nuclei by the shell-model calculations and, in cases like $^{110}$Te shown in the panel, it is the second $10^+$ state that is connected to the yrast $8^+$ state with strong E2 transition. As a result, the $B(E2,10_1^+\rightarrow8_1^+)$ value vanishes.
In the latter group, the $B(E2)$ values also show a large decrease after $I=4$ but to a extent that is much less than those in the fourth group, $^{126-132}$Te. 

To understand the behavior of the calculated $B(E2)$ values seen in Fig. \ref{SM4}, we notice that the ratio $E_{4^+}/E_{2^+}$ roughly equally to two for all known Te isotopes below $N=78$. But it decreases rapidly to around 1.2 in the semi-magic $^{134}$Te. Moreover, the ratio $E_{6^+}/E_{2^+}$ starts to decrease already at $N=72$, resulting in seniority-like spectra.
The seniority quantum number refers to the number of particles that are not paired to $J=0$. 
It is known that, for systems involving the same kind of particles, the low-lying states can be well described within the seniority scheme \cite{tal93}. This is related to the fact that the $T=1$ two-body matrix elements is dominated by monopole pairing interactions with $J=0$. 
The seniority coupling may be broken by the neutron-proton correlation if both protons and neutrons are present.
This indeed happens in the most neutron deficient Te isotopes close to $N=Z$, where the valence neutrons and protons are expected to occupy identical $g_{7/2}$ and $d_{5/2}$ orbitals and the neutron-proton correlation is expected to be strong. As a results, both the spectra and $E2$ transition properties show rather regular collective behaviors. On the other hand, for nuclei $^{126-132}$Te at the other end of the shell, the normal seniority coupling may prevail since the neutron-proton correlation involves particles in different shells and is much weaker. As a result, the energy gap between $6^+$ and $4^+$ as well as the $E2$ transition between the two states reduce significantly (e.g., $E2$ transitions between states with the same seniority is disfavored). The groups $N=58-64$ and $N=66-72$ fall between above two cases.

If the dimension is not too large, it is possible to project the wave function as a coupling of the proton group and neutron group with good angular momenta in the form $|\phi^p_{\pi}(J_{\pi})\otimes \phi^n_{v}(J_{v})\rangle$ where $J_{\pi}$ and $J_{v}$ denote the angular momenta of the proton group and neutron neutron group (see, e.g., \cite{Qi11,Xu2012}). 
The ground state for a even-even nucleus will be represented by a single configuration with $J_{\pi}=J_{v}=0$ if there is no neutron-proton correlation. The neutron-proton interaction induces contributions from configurations with higher angular momenta for the protons and neutrons as well as higher-lying configurations with the same angular momenta.
It is seen that, as expected, the $^{104}$Te ground state shows a high mixture of many component,
among which one has 47\% with $J_{\pi}=J_{v}=0$, 30\% with $J_{\pi}=J_{v}=2$, 12.5\% with $J_{\pi}=J_{v}=4$, and 6.7\% with $J_{\pi}=J_{v}=6$. 
For $^{106}$Te ground state the results are 41.7\% with $J_{\pi}=J_{v}=0$, 42 \% with $J_{\pi}=J_{v}=2$, 11.9\% with $J_{\pi}=J_{v}=4$.
For $^{106}$Te ground state the contribution from $J_{\pi}=J_{v}=0$ decreases further to 36.3\% while the contribution from $J_{\pi}=J_{v}=2$ increases to 46.6\%.
The wave functions for other low-lying states show a similar complex structure.
On the other hand, the wave functions for the low-lying states of the isotopes $^{126-132}$Te show a much simpler picture 
and are dominated by either neutron or proton excitations in many cases. The contributions from $J_{\pi}=J_{v}=0$ components are 
65.2\%, 74.4\% and 85.5\% for isotopes $^{128-132}$Te. The $2^+_1$ state in $^{132}$Te is dominated by $|\phi^p_{\pi}(J_{\pi}=2)\otimes \phi^n_{v}(J_{v}=0)\rangle$
whereas $6^+_1$ state is dominated by the proton excitation $|\phi^p_{\pi}(J_{\pi}=0)\otimes \phi^n_{v}(J_{v}=6)\rangle$ instead. The low-lying states for $^{128,130}$Te show a similar result.

It may be interesting to mention that a similar picture with rotational-like $B(E2)$ transitions and vibrational-like spectrum is also predicted for $N=Z$ nuclei like $^{92}$Pd \cite{Qi11} in relation to the quest for the possible existence of np pairing in $N\sim Z$ nuclei for which there is still no conclusive evidence after long and extensive studies (see, recent discussions in Refs. \cite{Ced11,Frauendorf201424,Qi11,Xu2012,Qi2015,Zhao20141,PhysRevC.89.014316,doi:10.1142/S0218301313300282}).
Moreover, the $\alpha$ formation amplitude may increase as a result of of the strong neutron-proton correlation.
There has already been a long effort answering the question whether the formation probabilities of neutron-deficient $N\sim Z$ Te and Xe isotopes are larger compared to those of other nuclei \cite{Seweryniak2006,Liddick2006}. 
We have evaluated within the shell-model approach the $\alpha$ formation amplitude \cite{Qi2015}. If the neutron-proton correlation is switched on, in particular if a large number of levels is included, there can be indeed significant enhancement of $\alpha$ formation amplitude. 

To summarize, we have done systematic calculations on the spectroscopy and transition properties of Te isotopes within the large-scale configuration interaction shell model approach. A monopole-optimized realistic interaction is used. The calculations reproduce well the excitation energies of the low-lying states as well as the regular and vibrational-like behavior of the yrast specta of $^{108-130}$Te (Fig. 1). The energies of the first $2^+$ and 
$4^+$ states as well as their ratios show rather a rather constant behavior when approaching $N=50$ in relation to the enhanced neutron-proton correlation (Fig. 2). On the other hand, a squeezed gap between the $6^+_1$ and $4^+_1$ states is expected when approaching $N=82$, resulting in seniority-like spectra. Those structure changes are also reflected in the calculated and available experimental E2 transition strengths.
The calculated $B(E2; 2^+_1\rightarrow0^+_1)$ show a parabolic behavior as a function of $N$, which is dominated by the contribution from the neutron excitation. Moreover, the calculations reproduced reasonably well the nearly constant behavior of the $B(E2)$ values of $^{114}$Te and $^{120-124}$Te along the yrast line (Figs. 3 \& 4). The anomalous constant behavior is related to the competition between the seniority coupling and the neutron-proton correlations. For neutron-deficient Te isotopes, the constant behavior of $B(E2)$ values can be extended to high spin values around $I=12,$ 14. Wheras for heavier isotopes, when the neutron-proton correlation gets weaker, the $B(E2)$ values can reduce significantly after $I=4$ and vanishes for the heaviest isotopes.

\section*{Acknowledgement}
This work is supported by the Swedish Research
Council (VR) under grant Nos. 621-2012-3805, and
621-2013-4323 and the G\"oran Gustafsson foundation.
We also thank D.S. Delion for discussions and for his efforts studying above-mentioned nuclei within the coherent state model and P. Maris for his effort calculating those nuclei with the code MFDn.
The
computations were performed on resources
provided by the Swedish National Infrastructure for Computing (SNIC)
at PDC, KTH, Stockholm.

\end{document}